# Optimum Drift Velocity for Single Molecule Fluorescence Bursts in Micro/Nano-Fluidic Channels

Lazar L. Kish [(1)], Jun Kameoka [(1)], Claes G. Granqvist [(2)], Laszlo B. Kish [(1)]

[(1)] *Department of Electrical and Computer Engineering, Texas A&M University, College Station, TX 77843-3128, USA*

[(2)] *Department of Engineering Sciences, The Ångström Laboratory, Uppsala University, P. O. Box 534, SE-75121 Uppsala, Sweden*

Photonic burst histograms can be used to identify single protein molecules in micro/nano-fluidic channels provided the width of the histogram is narrow. Photonic shot noise and residence time fluctuations, caused by longitudinal diffusion, are the major sources of the histogram width. This Communication is a sequel to an earlier Letter of ours [L. L. Kish et al., Appl. Phys. Lett. **99**, 143121 (2011)] and demonstrates that, for a given diffusion coefficient, an increase of the drift velocity enhances the relative shot noise and decreases the relative residence time fluctuations. This leads to an optimum drift velocity which minimizes the histogram width and maximizes the ability to identify single molecules, which is an important result for applications.

Single fluorescent color-based single-molecule detection [1-4] is the basis for a set of powerful modern techniques for sensing [1-4], including for air quality monitoring [5]. These techniques embrace micro- and nanofluidic devices, and a significant one is MAPS (microfluidic system for analyzing proteins in a single complex) [1]. MAPS utilizes quantum dots attached to proteins drifting in a micro/nano-fluidic channel. The quantum dots are excited by a laser beam and emit a group of photons (photon burst). Repeated experiments yield a histogram with a non-zero width of the photon number distribution [1].

We recently simulated the role of diffusion in single-molecule micro/nano-fluidics experiments and showed that diffusion adds extra fluctuations to the Poisson statistics (shot noise) of photon bursts [6]. Furthermore we demonstrated that the skewness and width of the distribution of photon numbers can be reconciled with residence time fluctuations in the excitation zone, analogously to the case of nanoparticle growth with log-normal size distribution in gas evaporation [7-9]. These fluctuations are caused by the diffusive (random walk) component of the molecular motion superimposed on the drift provided by the flow of the fluid and leading to log-normal-like distributions of residence time in the laser beam under a wide range of experimental conditions for which diffusion plays a decisive role [6]. The present Communication extends our analysis of the MEMS technique and demonstrates that there is an *optimum drift velocity to reach a minimum in the relative variance of photon burst size* (line width) because the effects of diffusion noise and shot noise scale in opposite directions for increased velocity.

We first use a simple qualitative one-dimensional model of the line width at varying drift velocities and assume that nothing but diffusion noise is present (no shot noise). The root-mean-square (RMS) displacement $\Delta$ of the diffusion during the residence time $t_{res}$ is given by the Einstein equation, i.e.,

$$\Delta = \sqrt{D t_{res}} , \qquad (1)$$

where $D$ is the diffusion coefficient of the molecule. For simplicity we assume the high-drift-velocity limit, which implies that the relative fluctuations of the residence time are small. Thus the mean residence time approaches the diffusion-free limit so that

$$\langle t_{res} \rangle = \frac{L}{V_d} , \qquad (2)$$

where $L$ is the diameter of the laser beam (taken to have uniform light intensity) and $V_d$ is the drift velocity. The RMS fluctuation of the residence time is proportional to the RMS displacement of the random walk component of the motion, meaning that the RMS fluctuation $\Delta t$ of the residence time is

$$\Delta t = \Delta / V_d . \tag{3}$$

Equations (1) to (3) then give

$$\Delta t = \sqrt{Dt_{res}} / V_d = \sqrt{DL/V_d} / V_d \propto V_d^{-1.5} . \tag{4}$$

In the large-photon-number limit (shot-noise-free limit), the number $N$ of emitted photons is proportional to the residence time, and the photon number fluctuation $\Delta N_D$ due to diffusion is proportional to the fluctuation of the residence time according to

$$N = \nu t_{res} , \tag{5}$$

$$\Delta N_D = \nu \Delta t_{res} , \tag{6}$$

where $\nu$ is the rate of photon emission. Equations (4) and (6) yield that the variance of the photon number fluctuation due to diffusion is

$$\left\langle \Delta N_D^2 \right\rangle = \nu^2 DL / V_d^3 \propto V_d^{-3} , \tag{7}$$

and Eqs. (5) and (7) demonstrate that the relative variance of the photon number fluctuation due to diffusion is

$$\frac{\left\langle \Delta N_D^2 \right\rangle}{N^2} = \frac{\nu^2 DL/V_d^3}{\nu^2 t_{res}^2} \approx \frac{DL/V_d^3}{t_{res}^2} = \frac{DL}{V_d} \propto V_d^{-1} . \tag{8}$$

The shot noise follows Poisson statistics [10], and thus the variance of the related photon number fluctuation $\left\langle \Delta N_s^2 \right\rangle$ is

$$\left\langle \Delta N_s^2 \right\rangle = N , \tag{9}$$

and the relative variance of the photon number fluctuation due to shot noise is

$$\frac{\left\langle \Delta N_s^2 \right\rangle}{N^2} = \frac{1}{N} = \frac{1}{\nu t_{res}} = \frac{V_d}{\nu L} \propto V_d , \tag{10}$$

where we supposed that nothing but shot noise is present and that diffusion noise is zero. Thus even though this simple model describes only the qualitative dependence of the relative variance of fluctuations at varying drift velocity, it is obvious that the two scaling exponents have opposite sign. Therefore, we may expect an optimum drift velocity corresponding to a minimum in the relative variance.

A minimum—i.e., an optimum drift velocity $V_{dm}$—will be located in the vicinity of the point where the two relative variances given by Eqs. (8) and (10) are equal, which leads to

$$\frac{V_{dm}}{\nu L} = \frac{\langle \Delta N_{sm}^2 \rangle}{N^2} = \frac{\langle \Delta N_{Dm}^2 \rangle}{N^2} = \frac{DL}{V_{dm}} \qquad (11)$$

and thus

$$V_{dm} \propto D^{0.5} \; . \qquad (12)$$

Two-dimensional computer simulations were performed as before [6] in order to illustrate the relationship between the variance of the fluctuations and the drift velocity. The molecule carrying the quantum dot drifted through a circular laser beam (with a diameter of 1700 steps and uniform intensity) via parallel paths with randomly selected distance from the center of the beam. While being in the beam, the quantum dot emitted photons with $\nu = 0.1$ photon/step rate. This situation in itself causes a widening of the photon burst distribution because molecules drifting along different paths spend different times in the beam. In addition to the drift velocity the molecules also executed a diffusion (random walk) process which introduced residence time fluctuations. The default widening of the photon burst distribution makes the minimum less sharp than expected from Eqs. (8) and (10) but Fig. 1 nevertheless gives clear evidence for the existence of a minimum in the relative variance.

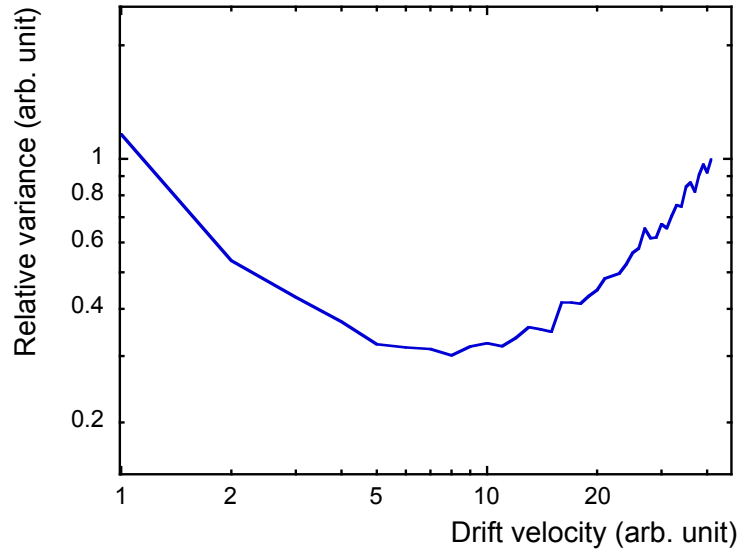

Figure 1. Relative variance of photon burst size (line width) versus drift velocity of a molecule in a micro/nano-fluidic channel.

Figure 2 shows the dependence of the optimum drift velocity, corresponding to the minimum in simulations such as the one leading to Fig. 1, on the diffusion coefficient. The optimum velocity appears to be a linear function of the square root of the diffusion coefficient, which is consistent with Eq. (12).

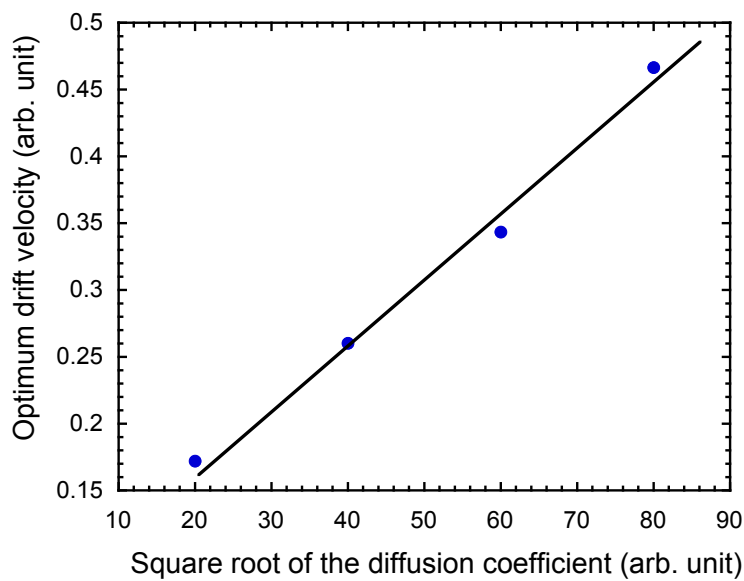

Figure 2. Optimum drift velocity versus the square-root of the diffusion coefficient.

In conclusion, simulations of fluorescent color-based single-molecule detection in micro/nano-fluidic devices have demonstrated that the molecular drift velocity can be optimized in order to minimize the inaccuracy in the measurement. The optimum drift velocity depends on the diffusion coefficient of the molecule, the size of the system and the intensity profile of the beam. This is an important result that can be used, among other things, to refine the novel MAPS technique.

Financial support was received from the European Research Council under the European Community's Seventh Framework Program (FP7/2007-2013)/ERC Grant Agreement 267234 ("GRINDOOR").